\def \s{~\rm{s}}
\def \km{~\rm{km}}

\def \K{~\rm{K}}

\def \erg{~\rm{erg}}

\def \yr{~\rm{yr}}
\def \pc{~\rm{pc}}
\def \kpc{~\rm{kpc}}

\documentclass[12pt,preprint]{aastex}

\usepackage{natbib}
\usepackage{epsfig}
\begin{document}

\title{THE SOURCE OF MASS ACCRETED BY THE CENTRAL BLACK HOLE IN
COOLING FLOW CLUSTERS}

\author{Noam Soker}
\affil{Department of Physics, Technion$-$Israel
Institute of Technology, Haifa 32000 Israel;
soker@physics.technion.ac.il.}

\begin{abstract}

This paper reports the study of the cold-feedback heating in cooling flow clusters.
In the cold-feedback model the mass accreted by the central black hole
originates in non-linear over-dense blobs of gas residing in
an extended region ($r \la 5-30 \kpc$); these blobs are originally hot, but then
cool faster than their environment and sink toward the center.
The intra-cluster medium (ICM) entropy profile must be shallow
for the blobs to reach the center as cold blobs.
I build a toy model to explore the role of the entropy profile and the population
of dense blobs in the cold-feedback mechanism.
The mass accretion rate by the central black hole is determined by the cooling
time of the ICM, the entropy profile, and the presence of inhomogeneities.
The mass accretion rate determines the energy injected by the black hole
back to the ICM.
These active galactic nucleus (AGN) outbursts not only heat the ICM,
but also change the entropy profile in the cluster and cause inhomogeneities
that are the seeds of future dense blobs.
Therefore, in addition to the ICM temperature (or energy), the ICM entropy profile
and ICM inhomogeneities are also ingredients in the feedback mechanism.
\end{abstract}


\section{INTRODUCTION}

In recent years it has become clear that the intra-cluster medium (ICM)
in cooling flow (CF) clusters of galaxies and CF galaxies must be heated,
and the heating process should be stabilized by a feedback mechanism.
(For recent papers and more references see
Begelman \& Ruszkowski 2005; Buote et al. 2005; Donahue et al. 2005;
Ensslin \& Vogt 2006; Fabian et al. 2005; Heinz \& Churazov 2005;
Hoeft \& Bruggen 2004; Mathews et al. 2006; Nipoti \& Binney 2005;
Omma \& Binney 2004; Ostriker \& Ciotti 2005; Vernaleo \& Reynolds 2006;
Voit \& Donahue 2005; Peterson \& Fabian 2006; Pope et al. 2006;
Brighenti \& Mathews 2006; Croton et al. 2006.)
As reported in most of these papers, the heating is based on active galactic
nuclei (AGN) which do one or more of the following: launch jets into the ICM;
inflate bubbles; cause turbulence; excite sound waves; and drive shocks.
In some clusters weak shocks are detected, e.g., Perseus (Fabian et al. 2006)
and M87-Virgo (Forman et al. 2005), while in some cases strong shocks
are driven to large distances in clusters, e.g., MS0735.6+7421
(McNamara et a. 2005) and Hercules A (Nulsen et al. 2005).
There are two questions regarding the AGN activity: (1) How is mass being feed to
the central black hole (BH)?; and (2) How does feedback heating maintain stability?
These questions are addressed in this paper.

Although heating substantially reduces the mass cooling rate, it
seems it cannot completely suppress the CF (see review by Peterson
\& Fabian 2006). The CF is defined as the process where mass near
the cluster's center cools to low temperatures ($\la 10^4 \K$),
and an inflowing still-hot gas replaces the cooling mass. Hicks \&
Mushotzky (2005) argue that in $\sim 2/3$ of the CF clusters that
they examined star formation could be the major sink for the
cooling gas inferred from X-ray observations. In the CF cluster
A1068, for example, there is an indication for star formation at a
rate of $20-70 M_\odot \yr^{-1}$, about equal to the mass cooling
rate in the central $\sim 30 \kpc$ inferred from X-ray
observations (Wise et al. 2004; McNamara et al. 2004). In the CF
cluster A2597, both extreme-UV and X-ray observations indicate a
mass cooling rate of $\sim 100 M_\odot \yr^{-1}$, which is $\sim
0.2$ times the value quoted in the past based on ROSAT X-ray
observations (Morris \& Fabian 2005).
Bregman et al. (2006)
find gas cooling through a temperature of $10^{5.5} \K$ in the CF
clusters Abell 426 and Abell 1795, with mass cooling rates of
$\sim 30-40 M_\odot \yr^{-1}$, which is $\sim 10 \%$ of the
cooling rate suggested in the past.
In the CF cluster A2029,
Clarke et al. (2004) find a substantial amount of gas at a
temperature of $10^6 \K$; a CF model gives a mass cooling rate of
$\sim 50 M_\odot \yr^{-1}$. Salome \& Combes (2006) argue that the
cold molecular gas in Perseus has its origin in the CF in this
cluster. These observations suggest that in many clusters the CF
is not completely suppressed, but rather occurs at a much lower
rate, $\sim 5-20 \%$ of that in models with no heating.

The moderate CF model (Soker et al. 2001; Soker \& David 2003; Soker 2004)
is one of the CF models where the mass cooling rate is low.
The moderate CF model is different from many early proposed heating processes whose
aim was to completely prevent the CF in clusters of galaxies. The main ingredient
of the moderate CF model is that the effective age, i.e., the time period since
the last major disturbance of the ICM inside the cooling radius
$r_c \sim 100 \kpc$, is much shorter than the cluster age (e.g.,
Binney \& Tabor 1995; Binney 2004; Soker et al. 2001).
The cooling radius is defined as the place where the radiative cooling time
equals the cluster age.
Heating in the moderate CF model was originally proposed to be intermittent
with strong shocks (Soker et al. 2001), as observed in some cases
(Nulsen et al. 2005; McNamara et a. 2005), but the basic idea holds for
steady heating, or for heating in short intervals (Binney 2004).

Pizzolato \& Soker (2005; hereafter PS05) study the feedback between heating
and cooling of the ICM in the frame of the moderate CF model, adopting the
mechanism in which a central BH accretes mass and launches jets and/or winds.
PS05 propose that the feedback occurs within the entire cool inner
region ($r \sim 5 - 30 \kpc$), where nonlinear over-dense blobs of gas with
a density contrast $\rho/\rho_a \ga 2$ cool fast and are removed from the ICM
before experiencing the next major AGN heating event;
$\rho$ is the density of a dense blob and $\rho_a$ the ambient density.
If the entropy profile is shallow, some of these blobs cool and sink toward the
central BH hole, while others might form stars and cold molecular clouds.
This scenario, where the BH is accreting cold gas, is termed ''cold feedback.''
This accretion process is different from the commonly assumed accretion mode
in feedback models, where the BH accretes hot gas from its vicinity via a
Bondi-type accretion flow (e.g. Churazov et al. 2002; Nulsen 2004;
Omma \& Binney 2004; Chandran 2005; Croton et al. 2006).
Soker \& Pizzolato (2005) propose that a large fraction
of the gas cooling to low temperatures of $T < 10^4 \K$ in CF
clusters gains energy directly from the central BH, (i.e., the
accretion disk or jets close to the BH). Most of this cool gas is
accelerated to nonrelativistic high velocities, $v_j \simeq
10^3-10^4 \km \s^{-1}$, after flowing through, or close to, an
accretion disk around the central BH.
Although massive outflows were not found yet in jets from AGNs in
CF clusters, massive AGN-outflows are known to exist (e.g., Morganti et al.
2005).
{{{ These massive outflows contain enough energy to heat the outflowing
gas and the ICM it interacts with, but not the entire CF region.
For heating the entire CF region faster jets are required. }}}

PS05 find the conditions under which the dense blobs formed by perturbations might
cool to low temperatures ($T < 10^4 \K$) and feed the BH to be as follows.
(1) An overdense blob must be prevented from reaching an equilibrium position in
the ICM; therefore, it has to cool fast, and the density, or more general the entropy,
profile of the ambient gas should be shallow.
(2) Nonlinear perturbations are required; they might have been
chiefly formed by previous AGN activity.
(3) The cooling time of these nonlinear perturbations should be shorter
than several times the typical interval between successive AGN outbursts.
(4) Thermal conduction around the blobs should be suppressed in order not to
evaporate the blobs.
This paper elaborates on the results of PS05 by constructing
a simple toy model aim at discovering the nature of feedback heating
and answering in part the two questions posed in the first paragraph of the
Introduction regarding the source of the mass accreted by the BH.

\section{THE MODEL}
I construct a toy CF cluster model as follows.

(1) {\it Density profile.}
I assume the density profile to be of the form
\begin{equation}
\rho = \rho_0 (1+ R^2)^{-1},
\label{rho}
\end{equation}
where
\begin{equation}
R \equiv \frac {r}{r_s},
\label{rr}
\end{equation}
$r$ is the distance from the cluster center (radius),
$\rho_0$ is the gas density at the center, and $r_s$ is the radius where
the density falls to half its central value.
This profile is simple, yet it keeps the main properties of the density profile
in CF clusters$-$flat near the center and steep at
$r \ga r_s \simeq 3-30 \kpc$, depending on the cluster, e.g.,
$r_s \sim 5 \kpc$ in M87/Virgo (Ghizzardi et al. 2004),
$r_s \sim 20 \kpc$ in Hydra A (David et al. 2001),
$r_s \sim 25 \kpc$ in A2052 (Blanton et al. 2001),
and $r_s \sim 30 \kpc$ in Perseus (Schmidt et al. 2002).

(2) {\it Dense blobs.} I follow PS05 and assume that the BH accretes dense
cold blobs.
These are formed from density perturbations scattered in the CF region.
The density perturbations themselves can be a result of previous AGN outbursts,
which are capable of disrupting the ICM (e.g., O'Sullivan et al. 2005).
The falling of dense blobs through the ICM in CF clusters was
discussed before, e.g., by Fabian (2003), who showed that such falling blobs can
release gravitational energy and contribute to the heating of the ICM.
The cold feedback model is different in that most of the heating
comes from the accretion of these blobs by the central BH.
For simplicity I assume that the perturbation profile does not depend on the
distance from the cluster center $r$.
Namely, the fraction of mass residing in dense blobs cooling faster than
their environment is $\delta_0$. As we'll see later, $\delta_0$ is an ingredient
in the feedback process and it might depend on time.

(3) {\it Mass cooling rate.}
The mass cooling rate of dense blobs in a shell with an inner radius $r$
and an outer radius $r+dr$ is
\begin{equation}
d \dot M_b =\frac{4 \pi r^2 \rho dr}{\tau_{\rm cool}} \delta_0 ,
\label{ddotm}
\end{equation}
where the cooling time at radius $r$ is
\begin{equation}
\tau_{\rm cool} = \frac{5}{2} \frac{nkT}{\Lambda n_e n_p} \equiv K_c \frac{kT}{\rho}.
\label{tauc}
\end{equation}
The symbols have their usual meaning: $n_e$, $n_p$, and $n$ are the electron, proton,
and total number density, respectively, $\Lambda n_e n_p$ is the cooling rate per unit
volume, $k$ is the Boltzmann constant, $T$ is the temperature, and
the last equality defines $K_c(T)\equiv (5/2) n \rho/\Lambda n_e n_p$.
In the relevant temperatures for the inner regions of CF clusters
$T \sim 2-5 \times 10^{7} \K$, and the dependence of $K_c$ on $T$ is weak.
I therefore take $K_c$  to be a constant.

(4) {\it The role of entropy profile.}
I assume a simple phenomenological relation between the number of
dense blobs that cool and are then accreted by the BH and the total
number of dense blobs at radius $r$.
Following the results of PS05 (PS05 Fig. 9),
the condition for accretion is that the entropy profile be shallow.
Blobs that fall too fast don't have time to cool much faster than
the environment they fall through; they reach an inner radius where they
are in equilibrium, i.e., their entropy equals that
of their new environment.
As entropy profiles becomes steeper, more blobs will reach equilibrium
and not fall to the center.
{{{ This process can be worked out by using the results of PS05 only if
the density-contrast distribution of perturbations is known.
Since this has to be worked out by 3D numerical simulations of the feedback
process, it is beyond the scope of the present paper.
Instead, I {\it postulate} a simple relation between the entropy profile
and the the fraction $f$ of cooling blobs that fall all the way to the center }}}
\begin{equation}
f= 1-A \frac{d \ln K_s}{d \ln r},
\label{f1}
\end{equation}
where $K_s=kT n^{-2/3}$ is the entropy, $A\sim 1$ is a constant of the toy model,
and if equation (\ref{f1}) gives $f<0$ I set $f=0$.
For example, if the temperature is constant, then by equation (\ref{rho})
\begin{equation}
\frac{d \ln K_s}{d \ln r} =-\frac{2}{3} \frac{d \ln \rho}{d \ln r}=
\frac{4}{3} R^2 (1+ R^2)^{-1}.
\label{f2}
\end{equation}
In that case $d \ln K_s/d \ln r = 4/3$ for $r \gg r_s$, and I therefore take $A=3/4$, so
that by substituting equation (\ref{f2}) in equation (\ref{f1}) one finds for a constant
temperature
\begin{equation}
f_T = (1+R^2)^{-1}.
\label{ft}
\end{equation}

(5) {\it Total mass cooling rate.}
Combining equations (\ref{rho}) - (\ref{f1}) gives the BH mass accretion rate
per unit radius (radial length) of mass originates at radius $r$
\begin{equation}
\frac{d \dot M_{BH}}{dr} =
\frac{4 \pi \delta_0 \rho_0^2 r_s^2}{ K_c k T}
\frac {R^2}{(1+R^2)^2}f.
\label{dtotmbh}
\end{equation}

(6) {\it AGN power.}
Part of the cooling mass is accreted by the BH, this part is
parameterized by $\delta_0$;  the other part being
expelled back to the ICM (Soker \& Pizzolato 2005), or forms stars.
I take the energy deposited by the central BH into the cluster to be
\begin{equation}
L_{BH}=\eta c^2 \dot M_{BH},
\label{ljets}
\end{equation}
where $c$ is the light speed.

\section{HEATING AND COOLING}

\subsection{Cooling}
The power emitted (in the X-ray) by gas residing inside a radius
$R_x$ is given by
\begin{equation}
L_x (R_x) = \int_0^{R_x} {\Lambda n_e n_p} 4 \pi r^2 dr
\label{lx1}
\end{equation}
Substituting for the density from equation (\ref{rho}) gives
\begin{equation}
L_x (R_x) =
\frac{5}{2} \frac{4 \pi \rho_0^2}{\mu m_H K_c} r_s^3
\int_0^{R_x} \frac{R^2}{(1+R^2)^{2}}dR =
\frac{5}{2} \frac{4 \pi \rho_0^2}{\mu m_H K_c} r_s^3
\frac{1}{2}\left[\frac {-R_x}{1+R_x^2} + \tan^{-1} R_x \right],
\label{lx2}
\end{equation}
where $\mu m_H$ is the mean mass per particle.
The cooling rate per unit radius in the model behaves as
$d L_x/dR \propto R^2/(1+R^2)^2$; it increases from the center to
$R=1$, and then decreases.
This qualitatively matches observation in the relevant range
(see fig. 8 of Voigt \& Fabian 2004).
The cooling rate of the entire cluster is given by substituting
$R_x \gg 1$ in equation (\ref{lx2})
\begin{equation}
L_x ({\rm total}) =
\frac{5 \pi^2}{2} \frac{\rho_0^2 r_s^3}{\mu m_H K_c}
\label{lxtot}
\end{equation}

\subsection{Blobs Accretion Rate Per Unit Radial Length}

\begin{figure}
\includegraphics[width =150mm]{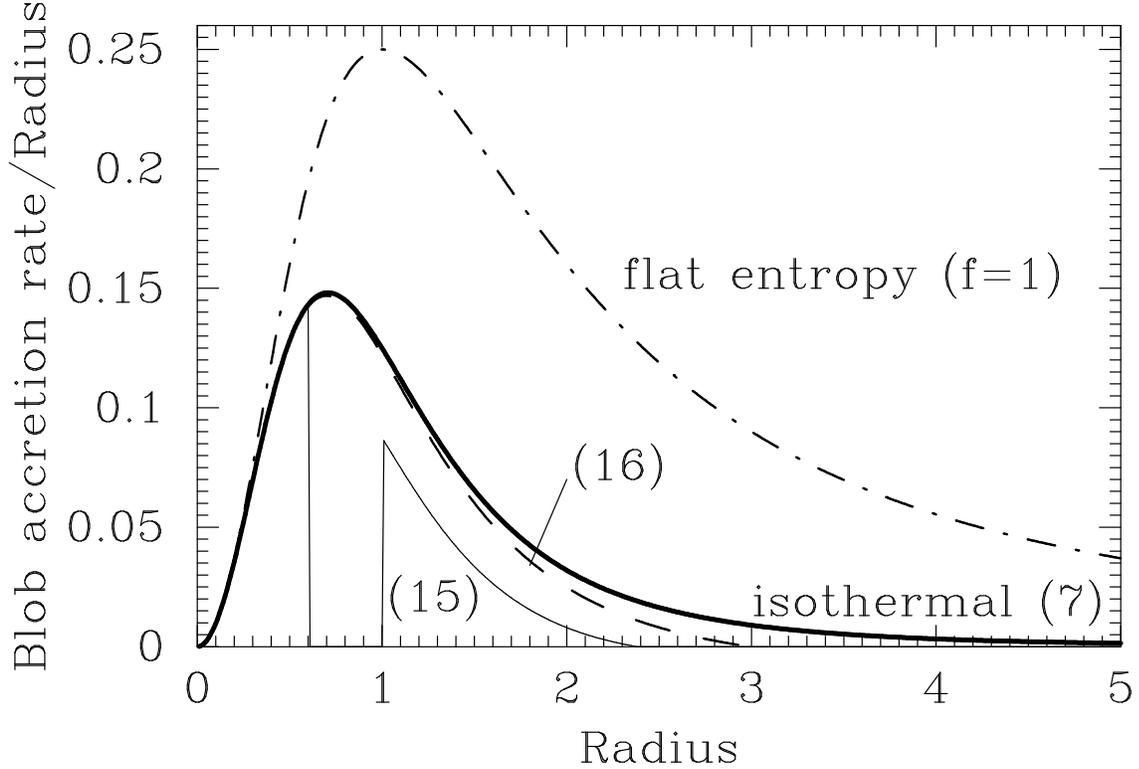}
\caption{The mass accretion rate by the BH of cold blobs per unit radial
length $R$ (${d \dot M_{BH}}/dR$) as given by equation (\ref{dtotmbh}).
The radius is $R \equiv r /r_s$, where $r_s$ is the radius where density falls
to half its value at the cluster center by equation (\ref{rho}).
${d \dot M_{BH}}/dR$ is plotted for different functions $f$ defined
in equation (\ref{f1}),
and in units of ${4 \pi \delta_0 \rho_0^2 r_s^3}/{ K_c k T}$.
The equation for $f$ for each line is given in parentheses.
Solid thick line: equation (\ref{dtotmbt}), namely, for the isothermal case
where $f=f_T$ (eq. \ref{ft}).
Dashed-dotted line: When the role of the entropy profile is neglected, $f=1$.
Solid thin line: The value of ${d \dot M_{BH}}/dR$ for the case where $f$ is
calculated from equation (\ref{fg}) by taking the temperature profile
from equation (\ref{t2052}).
Dashed line: As the solid thin line, but the temperature gradient
is given by equation (\ref{t87}). }
\label{feedf1}
\end{figure}

I consider now a case where the temperature is constant in the inner region.
Substituting the constant temperature case $f=f_T$ (eq. \ref{ft})
in equation (\ref{dtotmbh}) yields
\begin{equation}
\frac{d \dot M_{BH-T}}{dR} =
\frac{4 \pi \delta_0 \rho_0^2 r_s^3}{ K_c k T}
\frac {R^2}{(1+R^2)^3}.
\label{dtotmbt}
\end{equation}
The mass accretion rate in units of ${4 \pi \delta_0 \rho_0^2 r_s^3}/{ K_c k T}$
as given by equation (\ref{dtotmbt}) is plotted by the thick line
in Figure \ref{feedf1}.
For comparison the rate when the entropy role in the model
is not considered, namely $f=1$ in equation (\ref{dtotmbh}),
is plotted by a dashed-dotted line.
The entropy role in the toy model is to reduce the accretion rate
mainly from large distances.

I consider two other temperature profiles in calculating the factor $f$
according to
\begin{equation}
f_g = \frac {1}{(1+R^2)} -\frac{3}{4} \frac {d \ln T}{d \ln r},
\label{fg}
\end{equation}
in line with equation (\ref{ft}).
In the cluster A2052 (Blanton et al. 2001) the radius where the density
falls to half its central value is $r_s \sim 25 \kpc$,
and the temperature gradient can crudely be fitted by
\begin{equation}
\frac{d \ln T}{d \ln R} \simeq
 0 \quad (R \la 0.6);  \qquad
1.3  \quad (0.6  \la R \la 1); \qquad
0.2  \quad (1 \la R \la 4)
\label{t2052}
\end{equation}
%
%
where $R=1$ when $r=25 \kpc$ in this fit.
The temperature, and as a result, the entropy sharply rise
before density drops much.
This substantially changes the entropy profile at $R \ge 0.6$.
The mass accretion rate (per unit radial distance in units of
${4 \pi \delta_0 \rho_0^2 r_s^3}/{ K_c k T}$) of dense blobs when $f=f_g$
with the temperature given by equation (\ref{t2052}), is plotted
by the thin solid line in Figure \ref{feedf1}.
{{{ The contribution of blobs from regions beyond $R=0.6$ in the case
given by equation (\ref{t2052}) will be negligible, }}}
despite the non zero value at $R>1$.
{{{ Blobs can be formed in the region $R>1$, but the radiative cooling time
there is long, hence not many blobs will cool to low temperatures, and some
will reach equilibrium with their surroundings in the  region $0.6<R<1$.
In addition, as is discussed in section 3.3, in most cases the cold feedback
mechanism does not require accretion from $R>1$. }}}

At the other extreme, in some clusters the temperature stays constant
for large radii before rising.
In M87/Virgo (Ghizzardi 2004) the temperature gradient can crudely be fitted by
\begin{equation}
\frac{d \ln T}{d \ln R} \simeq 0.015 R^2 \qquad {\rm for} \quad \qquad R \la 4,
\label{t87}
\end{equation}
where $R=1$ when $r=5 \kpc$ in this fit for M87/Virgo.
The mass accretion rate (per unit radial distance in units of
${4 \pi \delta_0 \rho_0^2 r_s^3}/{ K_c k T}$) of dense blobs when $f=f_g$
with the temperature given by equation (\ref{t87}), is plotted
by the dashed line in Figure \ref{feedf1}.

The temperature and density profiles are shallow near the center for all clusters,
and as a consequence, the entropy profile is shallow there.
The entropy then rises sharply (Voit \& Donahue 2005; Donahue et al. 2006).
The temperature profiles used here represent the range of profiles
found in CF clusters.
In the majority of CF clusters, the entropy profile is almost flat near the
center; thus for addressing the questions posed at this study it
is sufficient to continue with the isothermal case.

\subsection{The Required Blobs Accretion Rate}

Equation (\ref{dtotmbt}) can be integrated to give the mass accretion rate originating
in radii smaller than $R_a$
\begin{equation}
\dot M_{BH-T} (R_a) = \int_0^{R_a} \frac{d \dot M_{BH}}{dR} dR  =
\frac{4 \pi \delta_0 \rho_0^2 r_s^3}{ K_c k T}
\frac{1}{8}\left[ \frac {R_a^3-R_a}{(1+R_a^2)^2} + \tan^{-1} R_a \right].
\label{dmbh}
\end{equation}

\begin{figure}
\includegraphics[width =150mm]{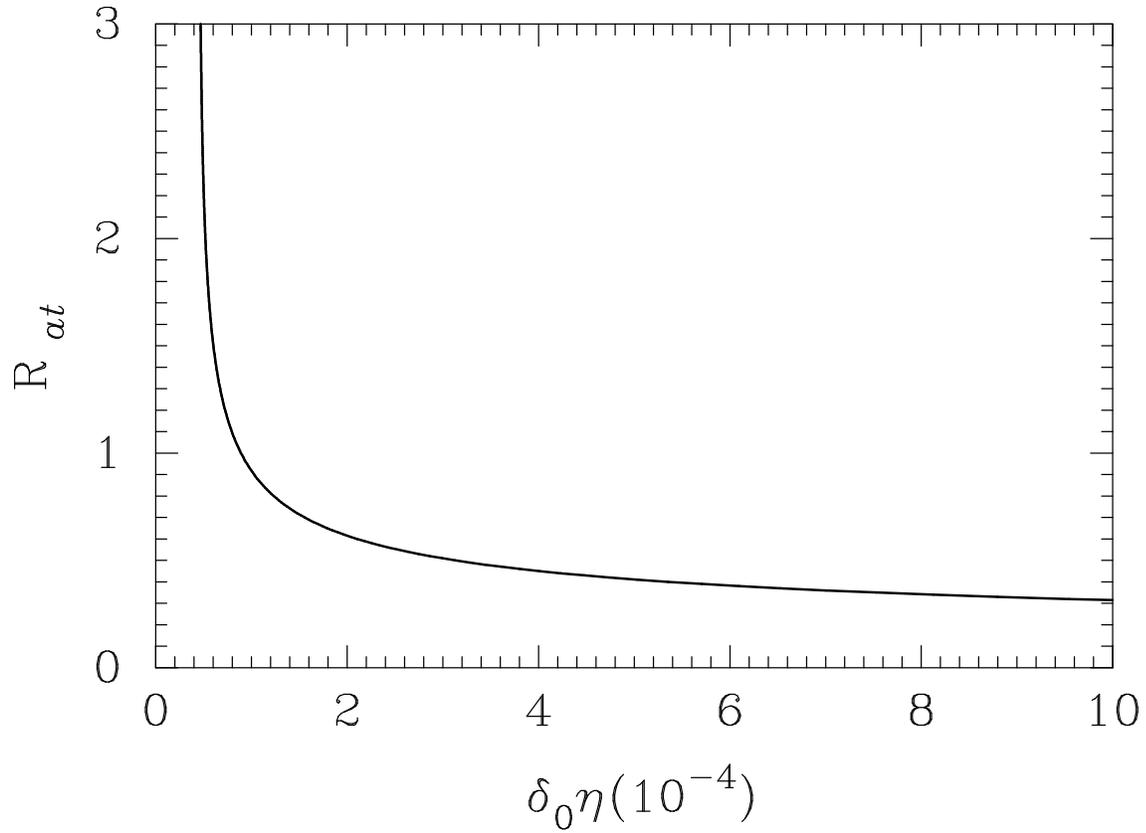}
\caption{To heat the entire ICM in the toy model, accretion
of dense blobs should occur from an extended region up to radius $R_{at}$.
The value of $R_{at}$ is plotted as a function of the efficiency factor
$\delta_0 \eta$ as given by equation (\ref{rat});
$\delta_0$ is defined in equation (\ref{ddotm}) and $\eta$ is
defined in equation (\ref{ljets}). }
\label{feedf2}
\end{figure}

To compensate for cooling in the entire cluster, accretion should occur
up to radius $R_a=R_{at}$ which is found by equating $L_x ({\rm total})$
(eq. \ref{lxtot}) with $L_{BH}$ (eq.  \ref{ljets}), where the mass accretion
rate from equation (\ref{dmbh}) is used in equation (\ref{ljets}).
This yields
\begin{equation}
\left[ \frac {R_{at}^3-R_{at}}{(1+R_{at}^2)^2} + \tan^{-1} R_{at} \right]=
\frac{3 \pi C_s^2}{\delta_0 \eta c^2}=
0.7 \left( \frac{T}{3 \times 10 ^7 \K} \right)
\left( \frac{\delta_0 \eta}{10^{-4}} \right)^{-1},
\label{rat}
\end{equation}
where $C_s = (5kT/3\mu m_H)^{1/2} \simeq 820 (T/3 \times 10 ^7 \K)^{1/2} \km \s^{-1}$
is the ICM speed of sound.
The scaling was chosen for a case where a fraction of $\delta_0=0.01$ of the mass is
in dense blobs that reaches the BH, and the energy deposited by the BH back to the
ICM is a fraction of $\eta=0.01$ of the rest energy of the accreted mass.
In Figure \ref{feedf2} the value of the radius $R_{at}=r_{at}/r_s$
as function ${\delta_0 \eta}$ according to equation (\ref{rat}) is plotted.

\section{THE FEEDBACK MECHANISM}

In Sections 2 and 3, the roles of the ICM entropy profile and the population
of dense blobs were explored by a toy model.
A cold blob can fall to the center of the cluster and be accreted by the
central BH if the entropy profile near its origin is not too steep (PS05).
If the entropy profile is steep, many sinking blobs will reach a region
where the ambient entropy is equal to their entropy, and they stop sinking.
A shallow entropy profile is a factor increasing the mass accretion rate,
{{{ although the radiative cooling time is still a more influential
factor (a shorter radiative cooling time increases the effective value of
$\delta_0$ in the model; see below). }}}
On the other hand, the entropy profile is influenced by AGN heating,
which itself results from mass accretion.
AGN heating of only the inner region makes entropy difference between the
still cooling outer region and the heated inner region smaller (a shallower
entropy profile).
In the next outburst cycle, the accretion rate will be higher because of
dense blobs falling in from regions further out.
AGN heating of the entire ICM, on the other hand, makes the entropy profile in
the $R \sim 1$ regions steeper.
The conclusion is that the entropy profile is an ingredient in the feedback process,
determining mainly the accretion rate from the region where most blobs reside in $R \sim 1$.
The role of the entropy profile was parameterized by $f$ (eq. \ref{f1});
{{{ this formula is an {\it ansatz} of the toy model. }}}

The dense cold blobs that are accreted by the central BH are formed in the
inhomogeneous ICM. The inhomogeneities in the ICM result from previous AGN activities
which form shocks, bubbles, and lift dense gas outward.
This implies that the presence of inhomogeneities in the ICM is also an
ingredient in the cold-feedback mechanism.
The role of the inhomogeneous ICM was parameterized in the toy model by the
factor $\delta_0$ (eq. \ref{ddotm}).

The results of Section 3 suggest the following feedback mechanism
in the frame of the cold feedback scenario.
For demonstrative purposes, the complicated feedback cycle is simplified
by considering three basic types of time periods.

{\it (P1) Between outbursts.}
The gas inside the CF region cools, with cold and dense
blobs cooling first. They sink toward the center.
Some of the cooling mass may form stars and cold clouds that are
not accreted by the central BH.
To reach the central BH, the entropy difference between the birth place
of the blob and the center must be small. If not, the dense blob will
reach a radius at which its entropy equals that of its surroundings.

{\it (P2) Accretion with a weak outburst.}
If the entropy profile is shallow only in the very inner region, $R \la 0.3$,
then in accordance with equation (\ref{dtotmbh}) blobs primarily from this inner
region are accreted; according to Figure \ref{feedf1} only a relatively small amount
of mass is accreted.
The AGN outburst is weak (depending on the efficiency factor ${\delta_0 \eta}$)
heating and disturbing only the very inner region.
The outburst has three effects: (1) Increasing the cooling time of the inner
region, thereby reducing the mass cooling rate from that region;
(2) Increasing the entropy in the inner region, making the
entropy profile shallower, allowing more dense blobs to be accreted
from regions further out in the next cycle; and (3) Disturbing the inner region,
forming density perturbations that are the seeds for dense blobs that
will be accreted in the next cycle.

{\it (P3) Strong outburst.}
After a weak AGN outburst, or several weak AGN outbursts, the entropy profile must
become shallower because the region near $R \sim 1$ cools and its
entropy decreases, while the heating of the inner region increases
the entropy in the inner region.
Because it was not heated efficiently by the weak outbursts, the region
at $R \sim 1$ is cooler and its entropy is lower than previously.
Both lower temperature and entropy ensure that more dense blobs from this
region are accreted by the central BH.
As seen from Figure \ref{feedf1} the region $R \sim 1$ could contribute
large amount of mass.
Figure \ref{feedf2} illustrates that if the perturbation fraction $\delta_0$
(eq. \ref{ddotm}) and the heating efficiency $\eta$ (eq. \ref{ljets}) are
such that $\delta_0 \eta \ga 10^{-4}$, then the accretion from the $R \sim 1$
region can heat the ICM in the entire cluster.
The strong heating of the region up to the cooling radius and beyond
(which extends to $R \gg 1$ in the toy model), increases the temperature and the cooling time.
This reduce the mass cooling rate, and we are back to  period P1.

The following comments are informative:
\newline
(1) The AGN outburst in CF clusters populate a continuous outburst energy
distribution. It is only for demonstrative
purposes that they were classified here `weak' or `strong'.
\newline
(2) If, in the toy model, the efficiency coefficient is
$\delta_0 \eta \la 5 \times 10^{-5}$,
then it is not possible to heat the entire cluster (Fig. \ref{feedf1}).
The outer region ($ R \sim 1-3$) cools, the entropy profile becomes shallower,
and the cooling time shorter. This implies that dense blobs with even smaller density
contrast can be accreted by the BH.
This results in a larger value of $\delta_0$, and a stronger feedback heating.
For the model to work, the value $\delta_0 \eta$ must be able to increase to
values $\ga 10^{-4}$ as cooling proceeds.
The population of dense blobs, as parameterized by $\delta_0$, is part of the
feedback heating mechanism.
This is an important additional ingredient to the assumed constant value
of $\delta_0$ in the toy model.
\newline
{{{ (3) The parameter $\delta_0$ stands for the fraction of cold blobs
that may cool to low temperature and be accreted by the BH,
according to equation (\ref{ddotm}),
which shows that the accretion rate is inversely proportional to
the gas cooling time $\tau_{\rm cool}$.
Practically, the dependence of the mass accretion rate on the cooling time
is stronger because as the cooling time get shorter more blobs can cool
to low temperature and be accreted (PS05).
Namely, the value of $\delta_0$ increases.
Therefore, the ICM cooling time is still the main ingredient of the
feedback model. }}}
\newline
(4) The entropy profile is also a significant ingredient in the feedback heating mechanism.
Not only the cooling rate, but also the entropy profile, determine the
mass accretion rate onto the central BH, and consequently the heating rate.
\newline
(5) This cold-feedback cycle is based on the existence of a CF,
but a moderate one (see Sec. 1).
Parcels of gas cool to low temperatures, but the mass cooling rate
is much smaller than in old CF models (Fabian 1994).
\newline
{{{ (6) A strong outburst can in principle lead to a turbulent inner region,
which can cause the entropy profile to stay shallow. However, this will not increase
the mass accretion rate because the entropy profile in the model is still
secondary to the radiative cooling time ingredient. Therefore, strong outbursts
even if flatten the entropy profile, will at the same time increase the radiative
cooling time, such that the mass accretion rate of blobs decreases. }}}

\section{COMPARISON WITH BONDI ACCRETION}
{{{
In a recent paper Allen et al. (2006; hereafter ADFTR) examined nine
elliptical galaxy and found a correlation between the Bondi accretion power
and the jet power $P_{\rm jet}$.
The Bondi accretion power is given by
\begin{equation}
P_{\rm Bondi} = \eta c^2 \dot M_{\rm Bondi}
\label{pbondi1}
\end{equation}
where $\dot M_{\rm Bondi}$ is the Bondi mass accretion rate.
ADFTR take $\eta =0.1$ and find the correlation
\begin{equation}
\log \left( \frac {P_{\rm Bondi}}{10^{43} \erg \s^{-1}} \right)
=0.62+0.77 \log \left( \frac {P_{\rm jet}}{10^{43} \erg \s^{-1}} \right).
\label{corr1}
\end{equation}
ADFTR use this correlation to argue that the accretion process is the Bondi
accretion process. I find nothing wrong in the results and interpretation
by ADFTR. However, I think their results are not in contradiction with the
results presented here, as I now explain.

{\it 1. Insufficient heating in clusters.}
As noted by ADFTR their results cannot be extrapolated to
cooling flows in cluster of galaxies.
To halt the cooling of gas in large clusters of galaxies the jet power should
be up to ${P_{\rm jet}} \sim 10^{45} \erg \s^{-1}$ (e.g., Birzan et al. 2004).
In this case the correlation found by ADFTR would give a jet power larger
than the power the accreted mass can supply.
As noted by ADFTR, cooling is expected in these clusters.
We suggest that part of this cooling is in blobs that further feed the
central BH.

{\it 3. Dependence on cooling time and entropy.}
The Bondi accretion rate is given by
\begin{equation}
\dot M_{\rm Bondi} = 4 \pi \lambda  (G M_{BH})^2 C_s^{-3} \rho.
\label{mbondi}
\end{equation}
Substituting the dependence of the speed of sound on temperature, and substituting
the Bondi accretion rate in the Bondi accretion power, we find
\begin{equation}
P_{\rm Bondi} = K_p^{\prime \prime} \left( \frac{\rho}{T^{3/2}} \right) M_{BH}^2,
\label{pbondi2}
\end{equation}
where $K_p^{\prime \prime}$, $K_p^{\prime}$, and $K_p$ are constants that
will be used in this section.
We note that for the temperature range of the X-ray emitting gas in
elliptical galaxies the cooling function dependance on temperature
can be approximated by $\Lambda \propto T^{-1/2}$, such that the cooling
time as given by equation(\ref{tauc}) is $\tau_{\rm cool} \propto T^{3/2}/\rho$.
The entropy dependance on temperature and density is
$K_s \propto  (T^{3/2}/\rho)^{2/3}$.
Using the cooling time and entropy as expressed above we can express the
Bondi accretion rate as
\begin{equation}
P_{\rm Bondi} = K_p^\prime (\tau_{\rm cool})^{\beta-1} (K_s)^{-3\beta/2}
M_{BH}^2,
\label{pbondi3}
\end{equation}
where $\beta$ is a parameter assumed to be $0<\beta<1$.
Substituting this relation in the correlation found by ADFTR yields
\begin{equation}
P_{\rm jet} =  K_p   (\tau_{\rm cool})^{1.3\beta-1.3} (K_s)^{-3.9\beta/2}
M_{BH}^{2.6}.
\label{corr2}
\end{equation}
This expression shows that the results of ADFTR can be interpreted as
the dependance of the jet power on the inverse of the entropy and cooling
time, as expected in the cold feedback mechanism.
For example,
for $\beta=0$ one finds $P_{\rm jet} \propto (\tau_{\rm cool})^{-1.3}$,
while
for $\beta=1$ one finds $P_{\rm jet} \propto (K_s)^{-2}$.

{\it 4. Dependence on mass.}
The relation between the bulge and BH mass is $M_{BH} \propto M_{\rm bulge}^{\kappa}$,
where $\kappa \sim 1.2-1.5$ (e.g., Laor 2001; Stuart \& Wyithe 2006 who find a more
complicated dependence).
The X-ray luminosity and mass of elliptical galaxies are positively correlated
(O'Sullivan et al. 2003); a positive correlation, although of a different kind,
holds in clusters of galaxies  (e.g., Yee \& Ellingson 2003).
The dependance of the jet power on the mass of the BH as found by ADFTR,
therefore, can be interpreted by a dependance on the galaxy mass or on
its X-ray luminosity as well.

{\it 5. Angular momentum.}
In M87 (Virgo), one of the galaxies studied by ADFTR, there is a large
gaseous disk of a radius of $\sim 35 \kpc$ (Macchetto et al. 1997).
Comparison with the Bondi accretion radius of $\sim 117 \pc$ (ADFTR),
implies that the accreted gas has a large specific angular momentum.
A large specific angular momentum in the gas might reduce substantially
the accretion rate (Proga \& Begelman 2003).

To summarize, I don't think the results of ADFTR undermines the model discussed
here. Basically, the results of ADFTR cannot be simply extrapolated from galaxies
to clusters of galaxies, and other possible interpretations of the correlation
they find are possible.
The cold feedback model is not in a stage where quantitative predictions can be made.
These will come in the future, with more sophisticated models.

The discussion here leads to a prediction of the proposed model.
The main predictions of the cold feedback mechanism were listed by PS05.
The basic prediction is that gas cooling to low temperatures ($<10^5 \K$)
should be detected at low to moderate levels in most CF clusters.
Also, the cold accreted gas will make the AGN activity similar in many
respects to AGN in spiral galaxies, where cold mass is accreted (PS05). From
the present results I add the prediction that the jet power in CF clusters
should correlate with the radiative cooling time of the ICM, its entropy,
and entropy gradient. In addition, the cluster mass and BH mass
can also play a role.
However, at present I cannot give any specific formula
for this dependence.
As was discussed above, the Bondi accretion power also correlates with
these same parameters.
To distinguish between the presently proposed role of the entropy profile
and other models, a comparison should be made between two CF clusters having
different entropy profiles, while all other properties are similar.
I predict that the cluster with shallower entropy profile will show
more cold gas near the center, a larger and more massive cold disk
near the BH, and a stronger AGN activity. }}}

\section{SUMMARY}

This paper discusses the cold-feedback mechanism in CF
clusters of galaxies (PS05).
In the cold-feedback mechanism, the feedback occurs with the entire cool
inner region ($r \la 5-30 \kpc$), where non-linear over-dense blobs of gas with a
large density contrast cool fast and are removed from the ICM
before experiencing the next major AGN heating event.
These blobs feed the BH at the center of the cluster.
This process implies the presence of a cooling flow (CF),
although a moderate one.

The feedback process was discussed in Section 4.
Results are now incorporated into the frame of the moderate CF model
by listing the ingredients of the feedback mechanism.
\begin{enumerate}
\item
{\it Energy Budget.} The feedback is basically a heating-cooling feedback:
cooling is primarily by radiative processes in the X-ray band;
heating is by AGN activity, mainly jets launched by an accreting massive BH.
Observations indicate that there is a substantial heating of the ICM, compensating
for most, {\it but not all}, of the energy radiated in the X-ray band.
The heating rate is determined to a large extent by the central BH
mass accretion rate.
The main factor that determines the mass accretion rate is the cooling
time of the ICM, which depends on the ICM density and on the ICM temperature.
The ICM temperature (or the ICM energy per unit mass),
is the main ingredient of the feedback mechanism.
The cold-feedback mechanism involves other ingredients in maintaining the feedback
stability.
The other ingredients help to prevent both running heating to
higher and higher temperatures and catastrophic cooling.
\item
{\it Dense cold blobs.} In the cold-feedback model the source of the accreted mass are
dense blobs resulting from inhomogeneities in the ICM. The ICM inhomogeneities result from
the same AGN activity that heat the ICM (an AGN outburst forms the dense blobs that
will be accreted in future cycles).
The role of the inhomogeneous ICM was modeled in the toy model by the factor $\delta_0$
(eq. \ref{ddotm}).
If, in some regions, there is no heating for a long period of time, such as in the
outer regions, the ICM cools with denser blobs cooling faster.
The long period of no heating ensures also that blobs with a smaller density contrast
have time to cool and increase their density contrast, resulting in a larger value of
$\delta_0$; namely an increase in the dense blobs population that might be accreted by the BH.
In the toy model heating efficiency is determined by the factor $\delta_0 \eta$,
where $\eta$ is the BH energy output efficiency (eq. \ref{ljets}).
For a reasonable range of $5 \times 10^{-5} \la \delta_0 \eta \la 10^{-3}$
accretion from the region $R \la 0.3-2$ can maintain the required accretion rate,
as demonstrated in Figure \ref{feedf2}.
\item
{\it Entropy profile.} If the entropy profile is steep, many sinking blobs will
reach a region where the ambient entropy equals their entropy,
and therefore they stop sinking.
Most relevant is the region $R \sim 1$, which can supply a large
amount of mass to the BH (Fig. \ref{feedf1}).
A weak AGN outburst, heating only of the regions $R \la 0.3$, makes the entropy
profile in the region $R \sim 1$ shallower, increasing the mass
accretion rate in the next cycle.
Heating the entire ICM, up to regions $R \gg 1$, makes the entropy
profile steeper in the region $ R \sim 1$.
The role of the entropy profile was parameterized by the function $f$ (eq. \ref{f1}).
\item
{\it Mass cycle.} Another possible, but not necessary, ingredient is a
slow (few$\times 1000 \km \s^{-1}$) massive outflow (Soker \& Pizzolato 2005).
This outflow starts in the accretion disk or its vicinity and carries a substantial
fraction of the cold ($ T \la 10^4 \K$) mass back to the ICM.
This gas is shocked and heated to high temperatures.
This outflow is significant to the mass cycle as well as to the energy cycle
(Soker \& Pizzolato 2005).
{{{ The energy is sufficient to heat the outflowing
gas and the ICM it interacts with, but not the entire CF region.
For heating the entire CF region faster jets are required. }}}
The mass cycle proposed by Soker \& Pizzolato (2005) is significantly different from
the circulation flow proposed by Mathews et al. (2004), because Mathews et al. (2004)
consider only hot gas with no cooling to temperatures below X-ray emission temperatures.
In contrast to the moderate CF model adopted here, Mathews et al. (2004) do not
consider the presence of a CF.
The recent mass and energy cycle model of Brighenti \& Mathews (2006) is more similar to
the model proposed by (Soker \& Pizzolato 2005) in that in contains some mass cooling
to low temperatures.

\end{enumerate}

The present results can be put in a more general perspective.
The new X-ray observations by the XMM-Newton and Chandra Telescopes show the ICM
in CF clusters to have {{{ a complex spatial distribution of temperature
and density, }}} and to posses an almost flat
entropy profiles near their centers.
{{{ the complex density distribution is assumed to exist also on an unresolved
small spatial scale, leading to the formation of cooling blobs. }}}
In the frame work of the cold-feedback scenario these observations support the presence
of CFs in these clusters, although moderate CFs.
Both  the detection of gas at temperatures of $\sim 10^6 \K$ and the detection of star
formation indicate the presence of CF in some clusters, but the more common
inhomogeneities and flat entropy profiles also argue for the presence of CFs.
The claim for a wide spread moderate CF phenomena is contrary to some claims,
made during the past six years since these two X-ray telescopes were launched,
supporting the absence of CFs.

\acknowledgments {{{ I thank Fabio Pizzolato for
helpful comments. }}}
This research was supported in part by the Asher
Fund for Space Research at the Technion.

\end{document}